\documentclass{article}

\usepackage{arxiv}

\usepackage[utf8]{inputenc} 
\usepackage[T1]{fontenc}    
\usepackage[bookmarks=false]{hyperref}       
\usepackage{url}            
\usepackage{booktabs}       
\usepackage{amsfonts}       
\usepackage{nicefrac}       
\usepackage{microtype}      
\usepackage{lipsum}
\usepackage{graphicx}
\usepackage{longtable}
\usepackage{glossaries}
\usepackage{enumerate}
\usepackage[export]{adjustbox}
\usepackage{amssymb}
\usepackage{pifont}
\usepackage{pbox}
\usepackage{algorithm}
\usepackage{algpseudocode}

\makeglossaries

\newacronym{COVID-19}{COVID-19}{Coronavirus Disease 2019}
\newacronym{CAM}{CAM}{class activation map}
\newacronym{CAP}{CAP}{community-acquired pneumonia}
\newacronym{CP}{CP}{consolidations in the periphery}
\newacronym{CQ}{CQ}{chloroquine}
\newacronym{CT}{CT}{computed tomography}
\newacronym{EHR}{EHR}{Electronic Health Record}
\newacronym{FC}{FC}{fully-connected}
\newacronym{FDA}{FDA}{fully-connected}
\newacronym{Grad-CAM}{Grad-CAM}{Gradient-weighted Class Activation Mapping}
\newacronym{HCQ}{HCQ}{hydroxychloroquine}
\newacronym{HPO}{HPO}{Hyper-parameters Optimization}
\newacronym{ICU}{ICU}{intensive care unit}
\newacronym{LIDC}{LIDC}{Lung Image Database Consortium}
\newacronym{MLL}{MLL}{Meta-level Learning}
\newacronym{NSAID}{NSAID}{non-steroidal anti-inflammatory drug}
\newacronym{PDP}{PDP}{Partial Dependence Plot}
\newacronym{ResNet}{ResNet}{Residual Network}
\newacronym{ROI}{ROI}{region of interest}
\newacronym{rRT-PCR}{rRT-PCR}{real-time reverse-transcriptase polymerase chain reaction}
\newacronym{SARS-CoV-2}{SARS-CoV-2}{Severe Acute Respiratory Syndrome Corona-virus 2}
\newacronym{WHO}{WHO}{World Health Organization}
\newacronym{GGO}{GGO}{ground-glass opacities}

\title{An Automated Approach for Timely Diagnosis and Prognosis of Coronavirus Disease}

\author{
 Abbas Raza Ali \\
  Faculty of Science and Technology\\
  Dept. of Computer Science, New York University\\
  Abu Dhabi, UAE \\
  \texttt{abbas.raza.ali@gmail.com} \\
  \And
 Marcin Budka \\
  Faculty of Science and Technology\\
  Bournemouth University\\
  United Kingdom \\
  \texttt{mbudka@bournemouth.ac.uk} \\
}

\begin{document}
\maketitle

\begin{abstract}
Since the outbreak of \gls{COVID-19}, most of the impacted patients have been diagnosed with high fever, dry cough, and soar throat leading to severe pneumonia. Hence, to date, the diagnosis of \gls{COVID-19} from lung imaging is proved to be a major evidence for early diagnosis of the disease. Although nucleic acid detection using \gls{rRT-PCR} remains a \emph{gold standard} for the detection of \gls{COVID-19}, the proposed approach focuses on the automated diagnosis and prognosis of the disease from a non-contrast chest \gls{CT} scan for timely diagnosis and triage of the patient. The prognosis covers the quantification and assessment of the disease to help hospitals with the management and planning of crucial resources, such as medical staff, ventilators and \glspl{ICU} capacity. The approach utilises deep learning techniques for automated quantification of the severity of \gls{COVID-19} disease via measuring the area of multiple rounded \gls{GGO} and \gls{CP} of the lungs and accumulating them to form a severity score. The severity of the disease can be correlated with the medicines prescribed during the triage to assess the effectiveness of the treatment. The proposed approach shows promising results where the classification model achieved 93\% accuracy on hold-out data. 
\end{abstract}


\glsresetall

\section{Introduction}
The \gls{COVID-19}, caused by \gls{SARS-CoV-2}, has proven to be highly infectious. The impact of the disease is such that the \gls{WHO} has declared it a global pandemic in March 2020, just over 2 months after the first cases have been reported in China~\cite{WHO_76_2020}. The exponential growth of \gls{COVID-19} cases raises a need for timely diagnosis of the disease to control its further spread. The nucleic acid detection using \gls{rRT-PCR} test is considered the \emph{gold standard} for the detection of \gls{COVID-19}~\cite{Ai2020}. However, studies show that this test is inadequate to effectively diagnose the disease due to several reasons including high-rate of false-negative indications~\cite{Rainer2020, Fang2020, Shi2020, Tahamtan2020, Gozes2020, Long2020, Basu2020, WangY2020, Ding2020}, limited availability of testing kits in several countries~\cite{He2020, Long2020, Basu2020}, and high-turnaround time of the test which normally takes from few hours up to two days~\cite{Liang2020, Fang2020, Shi2020, Tahamtan2020, Gozes2020, Long2020, Basu2020}. Thus, a negative \gls{rRT-PCR} result does not have to indicate a lack of \gls{COVID-19} infection, hence relying only on this test won't be sufficient for the accurate diagnosis of the disease~\cite{FDA2020, WangY2020}.

The constraints of \gls{rRT-PCR} can be comprehended by combining it with clinical examinations such as a non-contrast chest \gls{CT} scan. Chest \gls{CT} is regarded as evidence of a clinical diagnosis of \gls{COVID-19} which is playing a crucial role in diagnosing the disease at the early phase of constraining viral transmission~\cite{Shi2020}. Typical signs of the infection observed at the early stage from \gls{CT} scan are multiple rounded \gls{GGO} followed by pulmonary \gls{CP} of the lung at the later stage. In recent studies, a significant number of \gls{COVID-19} cases, about 8 in 10, are reported with mild symptoms~\cite{WHO_41_2020} where signs of infection include fever, nasal obstruction, runny nose, dry cough, sore throat, fatigue, respiratory symptoms, dyspnea, myalgia, and diarrhoea~\cite{WHO_76_2020}. Moreover, even asymptomatic patients hospitalised for observation showed lung abnormalities on a \gls{CT} scan, ranging from \gls{GGO} to \gls{CP}~\cite{long2020clinical}.

In order to address the discrepancies of \gls{rRT-PCR}, effective and efficient diagnosis and triage of \gls{COVID-19} are urgently needed. Therefore, an automated medical imaging-based approach to identify \gls{COVID-19} from the chest \gls{CT} scan has been proposed in this paper. We have achieved automated diagnosis of the disease using deep learning~\cite{LeCun2015} techniques, able to instantly identify radiographic changes from the \gls{CT} scans, allowing to make the diagnosis much earlier than the pathogenic testing like \gls{rRT-PCR}, which can, in turn, save critical time for minimising the spread of the virus~\cite{WangS2020}. In the proposed approach, once a patient is diagnosed as positive at an early stage, the same diagnostic process repeats periodically during the course of the treatment. Along with the diagnosis of the disease, an important aspect is the quantification of its impact. One way of achieving it using \gls{CT} is by computing the diameter of multiple patchy areas of \gls{GGO} and shadows of the \gls{CP} of the lungs~\cite{Zhu2019, ChenN2020}. This quantification of the disease serves two purposes, 1)~monitoring the progression of the disease over time, and 2)~in the absence of a vaccine, assessment of the treatment by correlating the effectiveness of the general medication prescribed to a patient. The medication includes cough syrup, \emph{ibuprofen}~\cite{Lesko1995, Pierce2010}, \emph{acetaminophen}~\cite{Pierce2010}, \glspl{NSAID}~\cite{Wen2017}, anti-inflammatory drug \emph{dexamethasone}~\cite{Johnson2020} and anti-malaria drugs, namely \gls{HCQ} and \gls{CQ} along with anti-viral medication \emph{remdesivir}. However, due to heartbeat irregularity side-effects associated with \gls{HCQ}, it is no longer prescribed to \gls{COVID-19} patients in some countries~\cite{FDA2020}.

The proposed approach revolves around an automated classification and measurement of \gls{COVID-19}. The high-precision has been achieved by training chest \gls{CT} segmentation and classification models using deep learning techniques. These techniques usually require large datasets to train (in excess of 1000 examples per class for a relatively simple classification task) that are not available particularly for \gls{COVID-19}. Therefore, transfer learning has been used to adapt a pre-trained network, which was originally trained on a large dataset - ImageNet~\cite{Russakovsky2015}, on a new domain having a comparatively small number of chest \gls{CT} images. Here, transfer learning is used in combination with \gls{MLL} to choose an optimal architecture, the approach suggested by ~\cite{Ali_Transfer2019}, and its \gls{HPO}~\cite{Ali_MetaRL2019} which could effectively adapt to a new domain. The deep learning techniques have been shown to work better if only critical, noise-free areas of the scan are fed for diagnosis~\cite{Shariaty2019}. Hence, before passing the image to the classification model, an existing deep learning-based image segmentation model of~\cite{Zhou2018} has been applied to the raw images as a pre-processing step to extract the interesting features~\cite{Adar2018, Dalca2019, ChenJ2020}. The \gls{COVID-19} positive classified cases are further quantified to measure the severity of the disease. To this end, the area of all the \gls{GGO} and \gls{CP} identified during classification are accumulated to compute an overall severity level. The classification and quantification processes repeat periodically until the patient is fully recovered and the severity score approaches zero. The severity is plotted on a timeline to help medical staff analyse the history of the patient along with the progression of the disease over time (see Figure~\ref{fig:progression} for an example). For the recovered patients, the timeline is correlated with the prescribed medications to get an indication of their effectiveness. This analysis also explains the patterns of patients' clinical data, such as age, gender, location, initial symptoms, and prior medical history versus the medications which leads to recovery. The novelty of this work is the development of an end-to-end pipeline for real-time diagnosis, quantification, progression analysis, and assessment of the medication to improve the overall treatment process of the \gls{COVID-19}. 

The remainder of this paper is organised as follows. Section~\ref{sec:Methodology} further elaborates the proposed approach. This section covers the datasets used to train and validate the chest \gls{CT} classification model. The disease quantification and assessment of treatment based on the severity of the diagnosis are also covered in this section. The results of various experiments have been presented in Section~\ref{sec:Results}. The paper is concluded in Sections~\ref{sec:Conclusions} and~\ref{sec:Future_work}.

\section{Methodology}
\label{sec:Methodology}
The proposed approach is capable of providing a timely diagnosis of \gls{COVID-19} along with quantification, progression, and assessment during the course of the treatment. These four phases are interlinked with each other where the process initiates with automated diagnosis of the suspected case. The non-contrast chest \gls{CT} is found to be the most effective clinical examination for the diagnosis of the disease since its outbreak~\cite{Liang2020}. In order to make the diagnosis efficient, a medical imaging-based model for the chest \gls{CT} was trained using deep learning techniques to accurately discriminate \gls{COVID-19} positive from negative cases. The classification results of each consecutive \gls{CT} scan have been recorded in the \gls{EHR} system along with the clinical information of a patient. The results include positive or negative classification of the disease and for positive cases, the measurement of the infected segments~\cite{Gozes2020}. 

\begin{figure}[htbp]
\centerline{\includegraphics[scale=0.90]{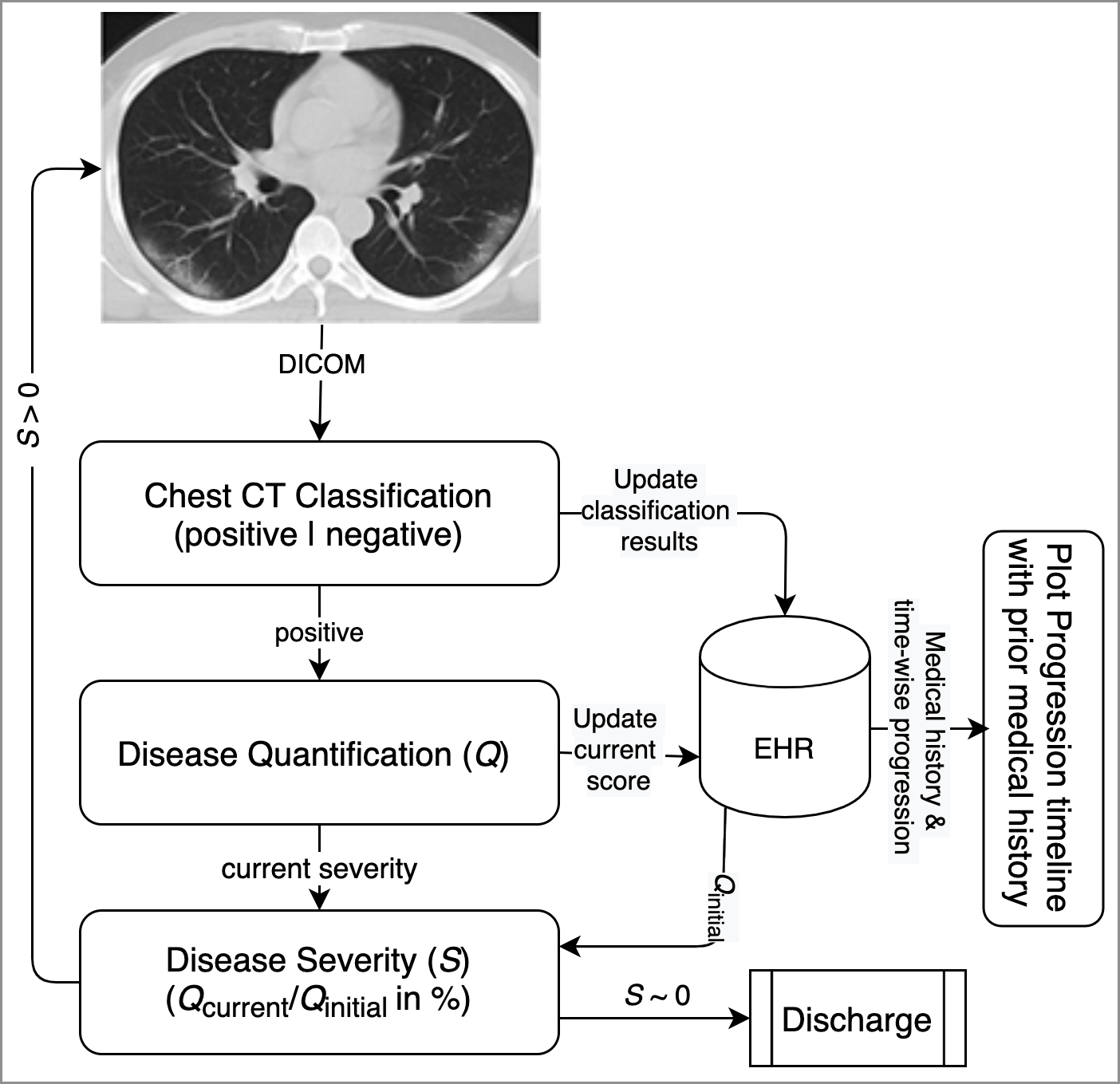}}
\caption{Early diagnosis and prognosis of the \gls{COVID-19} treatment workflow}
\label{fig:workflow}
\end{figure}

The initial diagnosis has been used for early detection of the disease which can be later cross-checked with the results of \gls{rRT-PCR}. Moreover, the early clinical diagnosis has proved effective for the patients showing at least mild symptoms which are mentioned earlier~\cite{LiangT2020}. Once a patient is confirmed with \gls{COVID-19} (using either \gls{rRT-PCR} or the approach proposed in this paper) and admitted for treatment, the monitoring process is initiated which periodically repeats the same diagnosis and quantification procedure to analyse the progression of the disease over the course of the treatment. The progression of the disease has been quantified by estimating the size of the abnormalities (\gls{GGO} and \gls{CP}) found during the diagnosis. The disease progression can be visualised over time since the beginning of the treatment to help medical staff assess the effectiveness of medication (Figure~\ref{fig:progression}). In the absence of the vaccine, this system can also correlate the progression of the disease with the prescribed medicines which are primarily not developed for \gls{COVID-19}. However, due to lack of access to relevant data, the correlation analysis has not been performed in this study. The workflow of the four-phase process is shown in Figure~\ref{fig:workflow}.

The following sections elaborate on the approach used to pre-process the training data, chest \gls{CT} modelling, quantification of the disease from the positive cases, and their assessment by correlating the severity level with the prescribed medicines. 

\subsection{Diagnosis} 

The chest \gls{CT} features generally observed in \gls{COVID-19} patients are bilateral multiple rounded \gls{GGO}, \gls{CP}, calcification, and lesion distribution~\cite{LiY2020, LiX2020, Liu2020, Jin2020}. In order to train a model that can discriminate scans pertaining these features from the normal ones, a dataset of \gls{COVID-19} patients' chest \gls{CT} scans were acquired from multiple sources which are listed in Table~\ref{tab:chestCT_datasources}. The merged dataset consists of 9,493 scans of 6,673 patients out of which 4,050 are \gls{COVID-19} positive, 2,274 have \gls{CAP}, and remaining 3,169 are healthy. 

\begin{table}[ht]
\caption{\label{tab:chestCT_datasources} List of sources from where chest \gls{CT} data has been acquired}
\centering
\begin{tabular}{c|r|r|r|r}
\hline
{\centering Source} & {\centering Patients} & {\centering COVID-19} & {\centering CAP} & {\centering Healthy} \\ \hline
\cite{Suny2020} & 1,525 & 1,495 & 1,027 & 0  \\ 
\cite{Ding2020} & 112 & 112 & 0 & 0  \\ 
\cite{Song2017} & 2,265 & 0 & 0 & 2,265 \\ 
\cite{Zhao2020} & 289 & 275 & 0 & 146  \\  
\cite{SoaresD2020} & 2,482 & 2,168 & 1,247 & 758 \\ \hline
Total & 6,673 & 4,050 & 2,274 & 3,169 \\ \hline
\end{tabular}
\end{table}


A total of 2,522 scans are collected from the Third Hospital of Jilin University, Ruijin Hospital of Shanghai Jiao Tong University, Tongji Hospital of Huazhong University of Science and Technology, Shanghai Public Health Clinical Center of Fudan University, and Hangzhou First Peoples Hospital of Zhejiang University~\cite{Suny2020}. The scans are further divided into 1,495 \gls{COVID-19} positive cases which are confirmed by \gls{rRT-PCR} and 1,027 scans belong to \gls{CAP} patients. The dataset used by~\cite{Ding2020} consists of only 112 \gls{COVID-19} cases as the study revolves around key findings of \gls{COVID-19}. \cite{Song2017} proposed classification of \gls{CAP} cases versus the healthy ones, hence, only healthy patients' scans of 2,265 have been collected from the \gls{LIDC}~\cite{LIDC}.

From \cite{Zhao2020} study, 275 \gls{CT} scans are obtained from 143 positive \gls{COVID-19} patients. The study combined this data with 146 healthy patients' scans. \cite{SoaresD2020} is a publicly available \gls{CT} scan dataset which is composed of 4,173 scans of 210 patients. The scans are further categorised into 2,168 scans, corresponding to 80 \gls{COVID-19} infected patients confirmed by \gls{rRT-PCR}, 1,247 scans of \gls{CAP} with 16 scans per patient on average and 758 healthy scans. The data is collected from the Public Hospital of the Government Employees and the Metropolitan Hospital of Lapa, both in Sao Paulo, Brazil. 

\begin{figure}[!htbp]
\centerline{\includegraphics[scale = 0.75]{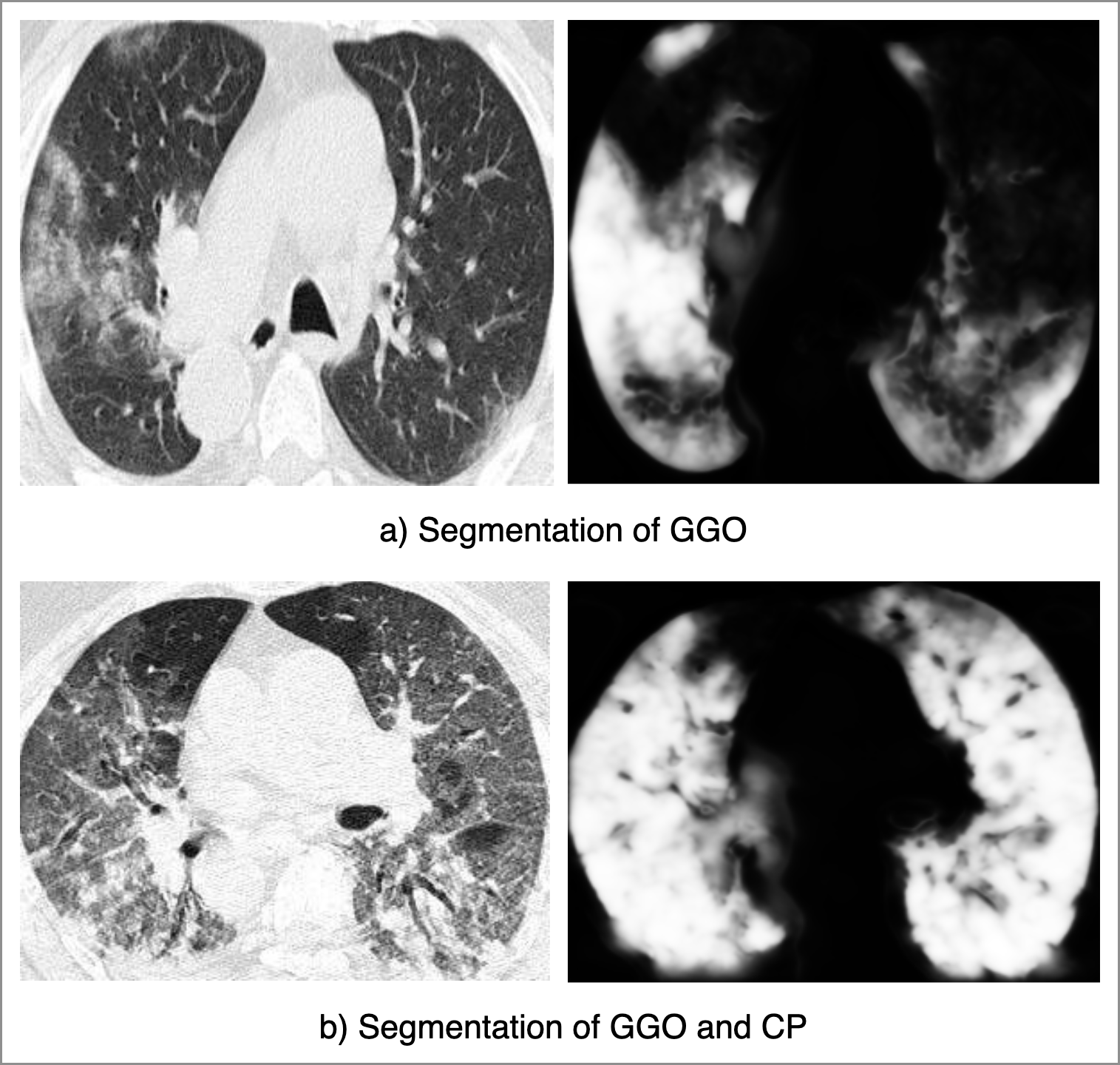}}
\caption{Segmentation of GGO and CP from lung CT}
\label{fig:segmentation}
\end{figure}

The scans and their labels from five different data sources were pre-processed, initially, by segregating the thorax region of both lungs from the background. These regions of a scan were segregated by applying a gray-level distribution of the Wiener-filtered image where different spikes of the distribution correspond to the lung, fat and muscle of the thorax region~\cite{Magdy2015}. The lung \gls{CT} scan is a combination of X-ray photons taken from different angles to produce cross-sectional slices of the lungs, therefore, arbitrarily two images per scan were selected. Further, automated semantic segmentation was applied to identify the infected \glspl{ROI}, such as, different organs and lesions, from the \gls{CT} slices which are crucial information for both diagnosis and quantification of the disease~\cite{Fan2020}. For this purpose, a commonly used medical image segmentation model, UNet++, trained on \gls{LIDC} by~\cite{Zhou2018}, was used to extract \glspl{ROI} from the image. 
The segmentation of \gls{GGO} at an early stage of the disease can be seen on the top whereas bottom scans show the segmentation of both \gls{GGO} and \gls{CT} at the critical stage of the disease as shown in Figure~\ref{fig:segmentation}.

\begin{figure}[!htbp]
\centerline{\includegraphics[scale = 0.9]{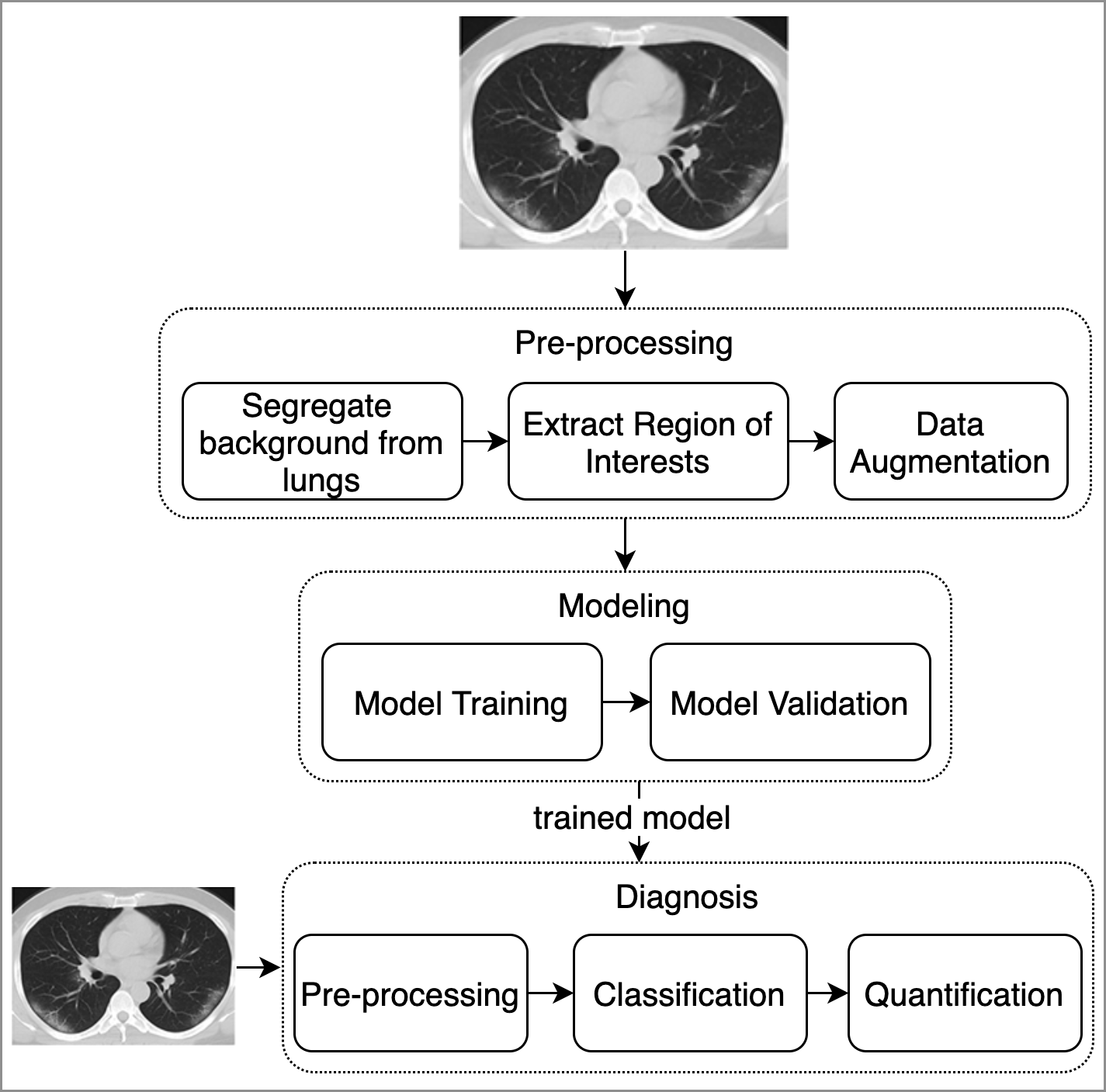}}
\caption{End-to-End data pre-processing, model training, and diagnosis pipeline}
\label{fig:pipeline}
\end{figure}

The data augmentation mitigates the issue of limited training data by adding synthesized image-label pairs into the training set. The pre-processed data was augmented with affine transformation which consists of 15 and 20 degrees rotation and translation respectively, random crop and horizontal flip~\cite{Zhao2020}. The scans gathered from multiple sources having different sizes were rescaled to 224x224 pixels and normalized using the per channel mean and standard deviation of the ImageNet dataset. An end-to-end data pre-processing, model training, and diagnosis pipeline are demonstrated in Figure~\ref{fig:pipeline}.

The training of deep learning models on a comparatively small dataset may lead to over-fitting where the model performs well on the training dataset, however, it doesn't generalise. In order to mitigate this limitation, transfer learning techniques have been adopted which acquire knowledge on a specific problem and reduce it to a different but related task~\cite{Yosinski2014}. Although, transfer learning is usually performed to fine-tune only the final \gls{FC} layer(s) of a pre-trained network to a new domain~\cite{Wang2017, Shin2016}, it does not guarantee state-of-the-art accuracy, particularly on relatively different tasks. Hence, in this work, a meta-transfer learning approach has been used for model training which fine-tunes an increasing number of layers of the network based on the complexity and relevance of both tasks as proposed by~\cite{Ali_Transfer2019}.

\begin{figure}[!htbp]
\centerline{\includegraphics[scale = 0.85]{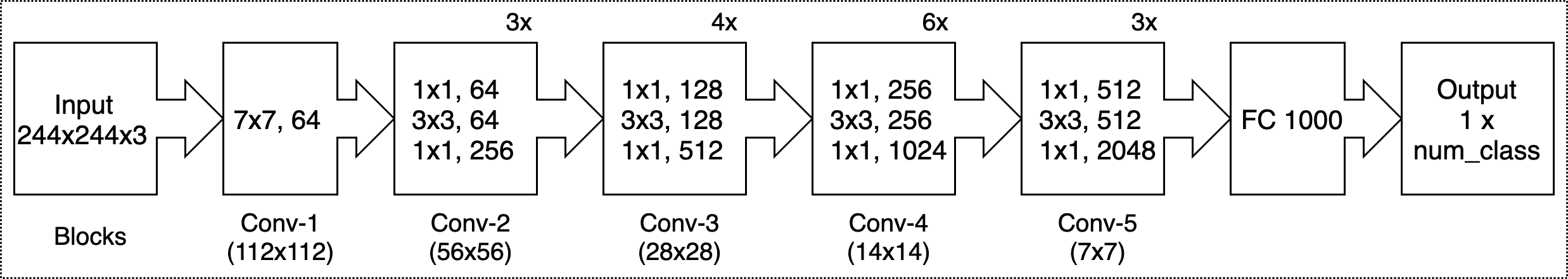}}
\caption{Schematic view of ResNet-50 architecture where all five convolutional blocks are re-trainable. The notation KxK, N in the convolutional block denotes a filter of size K with N channels. The number on the top of the block represents its repetition. The final layer predicts the number of output classes which is denoted by num\_class.}
\label{fig:resnet}
\end{figure}

For the disease classification modelling, a 50-layers \gls{ResNet}~\cite{He2016} was used with weights pre-trained on the ImageNet dataset. \gls{ResNet}-50 is composed of a set of residual blocks stacked on top of the input layer and followed by an \gls{FC} layer. A block consists of a sequence of two convolution layers with filter sizes 1x1 and 3x3, respectively, where a stride of 2 is used by the first convolution layer to reduce feature map size. \gls{ResNet}-50 provides a good trade-off between training time and performance. Generally, the first few layers of the network capture low-level features of the image like edges, curves, etc., where subsequent layers learn shapes and more abstract features. The final layers have learned more specific features corresponding to a particular problem in-hand. The pre-trained \gls{ResNet}-50 network was fine-tuned on \glspl{ROI} extracted from the \gls{CT} scans. The network was fine-tuned in an iteratively increasing number of blocks starting from the \gls{FC} layer, while the lower blocks of the network act as a fixed feature extractor. However, a block with its successive blocks was fine-tuned at a time, as elaborated in~\cite{Ali_Transfer2019}. A schematic view of \gls{ResNet}-50 architecture, labelled with the block numbers, can be seen in Figure~\ref{fig:resnet}.

The weights of the network were trained using Adam optimiser~\cite{Kingma2015} with a variable learning rate which decays by a factor of 10 on every alternative epoch with an initial value of 0.01. The training process automatically ends in case of no improvement found in the cross-validation loss for consecutive 3 epochs. The weights were initialised with Xavier-initialisation~\cite{Glorot2010} and a training batch-size of 16 was chosen. To classify a scan as \gls{COVID-19} positive or negative, the ratio of positive classified segments was compared against negative segments whereas the scan was declared positive when the ratio exceeds 0.5 threshold. The dataset was randomly divided into train, test, and validation sets with a ratio of 70\%, 15\%, and 15\% respectively. The \gls{CAP} category was excluded in the final training experiment. In order to thoroughly evaluate the proposed approach, a number of experiments have been carried-out by fine-tuning one layer/block per training cycle. Table~\ref{tab:data_training} summarises the number of positive and negative images in each set.


\begin{table}[!htbp]
~\caption{\label{tab:data_training} Dataset distribution into training, testing and validation sets}
~\centering
\begin{tabular}{l|r|r|r}
\hline
{\centering Partition} & {\centering COVID-19} & {\centering Healthy} & {\centering Total} \\ \hline
Training & 2,836 & 2,219 & 5,055  \\ 
Testing & 607 & 475 & 1,082  \\ 
Validation & 607 & 475 & 1,082  \\ \hline 
Total & 4,050 & 3,169 & 7,219  \\ \hline 
\end{tabular}
\end{table}

The network's hyper-parameters recommended by the Meta-RL~\cite{Ali_MetaRL2019} are listed in Table~\ref{tab:hpo}.

\begin{table}[!htbp]
~\caption{\label{tab:hpo} Hyper-parameter search space}
~\centering
\begin{tabular}{l|c}
\hline
{\centering Parameters} & {\centering Values (range)} \\ \hline
Dropout Rate & 0.0-0.5 \\  
Learning Rate & 0.01-0.00001 \\
Momentum & 0.6-0.99 \\ \hline
\end{tabular}
\end{table}

The next step of the proposed pipeline is the quantification of the disease. Not all patients diagnosed with positive \gls{COVID-19} will need intensive care. Accordingly,~\cite{NHC2020} classified \gls{COVID-19} into four severity levels including \textit{minimal}, \textit{common}, \textit{severe} and \textit{critical}. The \textit{minimal} and \textit{common} severity patients both show clinical symptoms however only in the \textit{common} severity level the chest \gls{CT} scan shows opacity in the lung. The \textit{severe} cases have signs of either respiratory distress, low blood oxygen saturation, or partial pressure of arterial blood oxygen~\cite{Yang2020}. The \textit{critical} patients are reported with organs and respiratory failure needing ventilation and intensive care. The \gls{CT} diagnosis and quantification of the disease could be very useful for patients falling under \textit{common} or \textit{severe} severity levels.

The severity score is the sum of the area of all the infectious segments that are found in the positively classified scans. Therefore, the gradient-weighed \glspl{CAM} is employed for detecting characteristic features from the image~\cite{Zhou2015}. The feature maps of the last convolution layer retain high enough spatial information which can be used for coarse localization of the pathological regions, whose size can, in turn, be related to the disease severity.


\begin{algorithm}  
\caption{Disease severity quantification algorithm}  \label{alg:algo1}
\begin{algorithmic}[1]
    \State $A^k \in \mathbb{R}^{h \times w}   \because k = 1, ..., N$ 
    \Comment{A is activation maps of the last convolutional layer, h and w are height and width of a feature map, and N represents total number of feature maps}
    \State $w^c = \frac{1}{h \times w} \sum_{i=1}^{w} \sum_{j=1}^{h} \frac{\emph{d}y^c}{\emph{d}A_{ij}}$ \Comment{weights of the class $c$ scores with respect to feature maps of \emph{A}}
    \State $\mathcal{Q} = \emph{ReLU}(\sum_{k=1}^{N} \max_{i=1}^{w} \max_{j=1}^{h} (w_k^c \times A_{ij}^k))$ \Comment{\emph{ReLU} is applied to features of positive influence on $c$}
\end{algorithmic}
\end{algorithm}

The disease severity quantification algorithm is outlined in Algorithm~\ref{alg:algo1} which has been computed for each segment of a scan for which the model has classified infected. The quantification value $\mathcal{Q}$ for all the segments of a scan has been summed up which is elaborated in Figure~\ref{fig:progression}.


\subsection{Assessment of the Treatment}
The assessment of the treatment is the post-processing phase of the approach which comes after diagnosis and quantification of disease. The quantification score is further used for patient-specific monitoring of the disease progression during the course of treatment. The progression is monitored by the severity score $\mathcal{S}$ which normalizes the quantification score of a particular day by the initial score, see Equation~\ref{eq:1}. The severity score is extrapolated to forecast the condition of the critical patients which allows the hospitals to plan ahead for the crucial resources like medical staff, ventilators and \glspl{ICU}.

\begin{equation} \label{eq:1}
    \mathcal{S} = \frac{\mathcal{Q}_{current}}{\mathcal{Q}_{initial}} \times 100
\end{equation}

\begin{figure}[!htbp]
\centerline{\includegraphics[scale = 0.50]{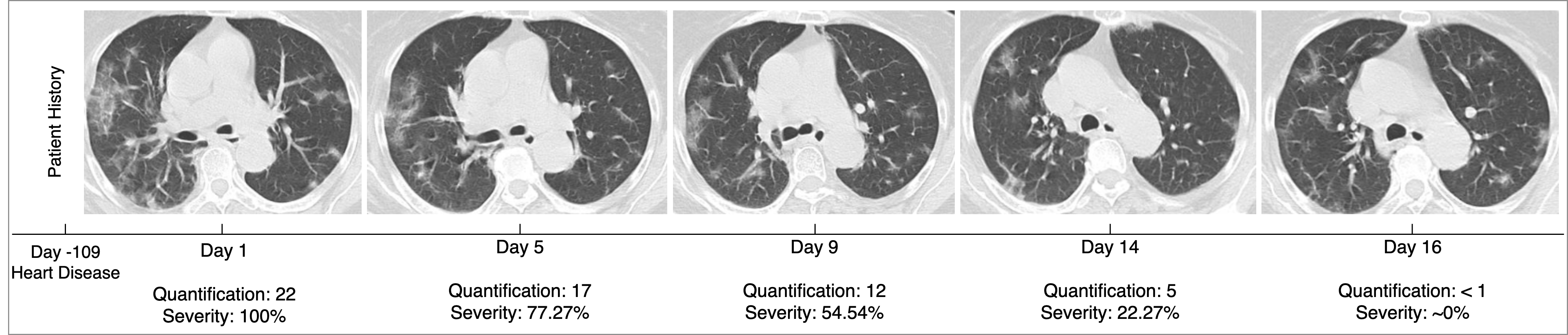}}
\caption{The \gls{COVID-19} progression analysis which quantifies the current severity along with relative progression of the disease}
\label{fig:progression}
\end{figure}

The disease progression is plotted over time since the beginning of treatment along with prior medical history and medication prescribed during the treatment. Since there is no vaccine available for \gls{COVID-19}, this approach correlates the progression of the disease with prescribed medicines. Such analysis can help physicians to assess the effectiveness of medication on patients and come up with a pattern of the patient's severity level and medication. \gls{FDA} granted permission to use some general medicines, mentioned earlier, for the treatment of severely affected \gls{COVID-19} patients. Those medicines are correlated with the progression to evaluate their effectiveness along with the clinical data and prior medical history of the patient to predict the most suitable set of medicines for the new patients~\cite{LiangW2020}. The \gls{COVID-19} prognostics has been demonstrated in Figure~\ref{fig:progression}.

\section{Experimental Results}
\label{sec:Results}
A comprehensive set of experiments has been conducted to evaluate the effectiveness of the proposed approach. These experiments evaluate various models which are fine-tuned on a different combination of the layers of a network. The experiments were performed on Nvidia Tesla v100 GPU. The disease classification is modelled using \gls{ResNet}-50 architecture. The architecture takes less time to train a model as compared to the other architectures such as Inception-v3~\cite{Szegedy2016} and Inception-ResNet-v2. The network was fine-tuned in an iteratively increasing number of blocks starting from the final block, only one block was fine-tuned in a training cycle at a time. 

\begin{table}[!htbp]
~\caption{\label{tab:accuracyTbl} The block/layer wise results of the \gls{COVID-19} classification model}
~\centering
\begin{tabular}{l|r|r|r|r}
\hline
{\centering Block/Layer} & {\centering Accuracy} & {\centering Precision} & {\centering Recall} & {\centering F1 Score} \\ \hline
FC     & 93.75 & 92.15 & 90.53 & 91.15 \\  
Conv-5 & 91.25 & 90.39 & 87.19 & 88.53 \\  
Conv-4 & 88.87 & 87.87 & 83.75 & 85.77 \\  
Conv-3 & 86.27 & 85.23 & 80.07 & 82.57 \\  
Conv-2 & 84.97 & 83.33 & 78.57 & 80.88 \\  
Conv-1 & 82.14 & 80.68 & 76.07 & 78.31 \\ \hline 
\end{tabular}
\end{table}

Table~\ref{tab:accuracyTbl} summarises the performance of the block-wise network fine-tuning on hold-out test-set. The performance metrics consist of accuracy, precision, recall, and F1 score. The fine-tuning of the final \gls{FC} layer gives the best classification with 93\% accuracy, 92\% precision, 90\% recall, and 91\% F1 score. These results were extracted from a training cycle consuming pre-processed segments of images. However, the fine-tuning of the rest of the blocks could outperform the current best accuracy in case of the availability of more training data. The precision and recall show promising results with a low false-negative rate.

\section{Conclusions} \label{sec:Conclusions}
This paper presents an effective and efficient pipeline for diagnosis and triage of a suspected patient which is crucial to control \gls{COVID-19} and to provide better treatment of the infected patients. The proposed approach uses deep learning-based \gls{CT} image classification techniques for the diagnosis of the disease. These techniques usually require large training datasets, however, the transfer learning technique was used to adapt a pre-trained model on the available \gls{CT} scan data. In fact, deep learning techniques are highly sensitive which requires focusing on critical areas of the \gls{CT} scan for precise diagnosis. Hence, a deep learning-based image segmentation technique was applied to the raw images as a pre-processing step to extract useful features from the image. The diagnosis of the disease from a \gls{CT} scan takes significantly less time than the \gls{rRT-PCR} test. Once a patient is admitted to the hospital for treatment based on the results of the diagnosis, the same process repeats periodically during the course of the treatment. Further, the deep learning classification model also quantifies the diseases by computing the area of \gls{GGO} and \gls{CP} which are identified by the model as infectious segments. This process helps physicians to monitor the progression of the disease over time from an early stage of the treatment. Moreover, in the absence of anti-viral medicine for \gls{COVID-19}, a post-processing analysis correlates the general medication prescribed to a patient with the quantification of the disease severity score for better assessment of the treatment. 

The novelty of this work is the development of an end-to-end pipeline for timely and accurate diagnosis, quantification, progression analysis, and assessment of treatment of \gls{COVID-19}. The effectiveness of the approach relies on the performance of the diagnosis both in terms of speed and accuracy. Hence, the Automatic Machine learning approach, such as \gls{MLL} has been used to find the optimal number of network blocks that need to be fine-tuned. The final model gave an accuracy of 93\% on the hold-out set with high precision and sensitivity.

\section{Future Directions} \label{sec:Future_work} 
The disease diagnosis process is composed of two deep learning models including segmentation and classification. These models have millions of parameters across multiple layers/blocks which makes the modelling process hard to explain and less transparent. Perhaps, only technical experts can interpret these models in the context of the transformations and operations that the data goes through. However, for the end-users, mostly high-level performance measures are the only parameter to trust in the working of the models. It would become critical when the end-user of these models belongs to the domain of medical science who needs to make a decision based on very abstract insights. Therefore, the prognosis of the disease can be further enhanced by interpreting the following areas of diagnosis:
\begin{enumerate}
    \item Data Audit: What if a model is overfitted due to class biases and gives high performance? A detailed data audit that covers the class distribution along with other information would be useful for the end-user.
    \item Model Interpretation: 
    \begin{enumerate}
        \item Along with the data audit, explainability of the black box deep learning models by demonstrating how a model processed a scan by interpreting each layer/block of the network?
        \item What kind of examples does the model perform poorly on? 
        \item Can a prediction be attributed to adversarial behavior, or to undesirable priors in the training set?
    \end{enumerate}
    \item Quantification of the disease: For a \gls{COVID-19} positive case, how critical is the condition of the patient. One possible solution could be precise quantification of the disease by volumetric measurement of multiple patchy areas of \gls{GGO} and shadows of the \gls{CP} of the lungs. 
    \item Localization of Disease: The model's outcome of \gls{COVID-19} is based on the patches which actually belong to \gls{COVID-19} or \gls{CAP}. This can be interpreted by visualizing the final feature maps of the network to discriminative regions of the scan. This can be achieved by localization of disease using \glspl{PDP}~\cite{Buhrmester2020}, \gls{Grad-CAM}~\cite{Selvaraju2019} and saliency mapping~\cite{Mundhenk2020}.
    \item Relating insights of scan and clinical data: What value patient's clinical data, demographics, oxygen saturation level, etc. can add along with the diagnosis of the scan towards explainability. A possibility is through the clustering of the clinical data and diagnosis to uncover insightful patterns.
\end{enumerate}

\section*{Acknowledgements}
The authors wish to thank Dr. Muhammad Naeem medical specialist at Services Hospital for taking part in the subject discussion and review of this work.

\bibliographystyle{unsrt}  
\bibliography{refs}  

\end{document}